\documentclass[runningheads]{llncs}
\usepackage[T1]{fontenc}
\usepackage{algorithmic}
\usepackage{algorithm}
\usepackage{array}
\usepackage{textcomp}
\usepackage{stfloats}
\usepackage{url}
\usepackage{verbatim}
\usepackage{cite}
\usepackage{comment}
\usepackage[svgnames]{xcolor}
\usepackage{tcolorbox}
\usepackage{array}
\usepackage{setspace}
\usepackage{cleveref}
\usepackage{subfig}
\usepackage{wrapfig}
\usepackage{hyphenat}
\usepackage{graphicx}
\definecolor{OliveGreen}{RGB}{183, 207, 178}
\definecolor{DarkYellow}{HTML}{DEA601}
\definecolor{DarkOrange}{HTML}{ED7D31}
\definecolor{DarkBlue}{HTML}{4472C4}

\crefname{section}{sec.}{sec.}

\newcommand{\supplement}{\url{https://shorturl.at/xEQR7}}

\begin{document}
\title{Seeing is Believing: The Role of Scatterplots in Recommender System Trust and Decision-Making}
\titlerunning{Seeing is Believing: Recommender System Trust}
\author{Bhavana Doppalapudi\inst{1} \and
Md Dilshadur Rahman\inst{2}   \and
Paul Rosen\inst{2}}
\authorrunning{B. Doppalapudi et al.}
\institute{University of South Florida, Tampa, FL 33620, USA
\email{bhavanadoppalapudi@gmail.com} \and
University of Utah, Salt Lake City, UT 84112, USA\\
\email{\{dilshadur,prosen\}@sci.utah.edu}}
\maketitle              %
\begin{abstract}
 The accuracy of recommender systems influences their trust and decision-making when using them. Providing additional information, such as visualizations, offers context that would otherwise be lacking. However, the role of visualizations in influencing trust and decisions with recommender systems is under-explored. To bridge this gap, we conducted a two-part human-subject experiment to investigate the impact of scatterplots on recommender system decisions. Our first study focuses on high-level decisions, such as selecting which recommender system to use. The second study focuses on low-level decisions, such as agreeing or disagreeing with a specific recommendation. Our results show scatterplots accompanied by higher levels of accuracy influence decisions and that participants tended to trust the recommendations more when scatterplots were accompanied by descriptive accuracy (e.g., \textit{high}, \textit{medium}, or \textit{low}) instead of numeric accuracy (e.g., \textit{90\%}). Furthermore, we observed scatterplots often assisted participants in validating their decisions. Based on the results, we believe that scatterplots and visualizations, in general, can aid in making informed decisions, validating decisions, and building trust in recommendation systems.

\keywords{Machine Learning  \and Visualizations \and Trust \and Decision-Making.}
\end{abstract}

\setstretch{1}

\section{Introduction}
\label{sec.intro}

Recommender systems are ubiquitous in diverse contexts, from low-risk activities like traffic prediction~\cite{kay2016ish} and movie suggestions~\cite{surendran2020movie} to high-risk scenarios like delivering child care~\cite{chouldechova2018case} and homeless services~\cite{kube2019allocating}. The widespread use of recommender systems highlights the need to understand the factors influencing users' trust in these systems. Significant research has been dedicated to studying and improving trust by enhancing transparency and providing information such as accuracy and explanations for recommendations~\cite{kizilcec2016much, yin2019understanding, kunkel2019let, lai2019human}. 

While providing accuracy offers insights into recommender system performance, research often overlooks the nuanced meaning of accuracy, which is context-specific (e.g., due to class imbalance, classifier type, i.e., binary vs.\ multiclass, etc.), and expecting the general population to possess this knowledge is unreasonable. 
A simple scenario is to consider a condition that is observed only 10\% of the time. A recommender system that always reports false would still exhibit a 90\% accuracy! While experts possess the knowledge to employ additional metrics to illustrate this imbalance, the same cannot be expected from the general population, 
highlighting the need to investigate the effects of providing additional context, such as visualizations, to assist people in decision-making.   

The use of visualizations in understanding people's trust in recommender systems has been relatively underexplored. Most of the current research focuses on interactive visualizations and tools~\cite{collaris2020explainexplore, ahn2019fairsight, ren2016squares} to explore, analyze, and validate the recommendations and performance of recommender systems. While offering tools for analysis and exploration proves advantageous for professionals and researchers, translating these to the general population seems unreasonable. Furthermore, there exists a lack of focus on how including simple visualizations that depict data and offer extra contextual information regarding the features utilized by the recommender system supports trust and decision-making. 

We describe the findings of a crowdsourced study on how the inclusion of visualization, scatterplots in particular, with accuracy, shapes people's decisions and trust in recommender systems. Our first experimental task requires participants to consider the presented scatterplots along with accuracy and make a decision on selecting 1 of the 2 provided recommender systems, with the selection of 1 over the other serving as a proxy for trust.  The second experimental task requires participants to consider the presented scatterplot and decide whether to agree or disagree with a recommendation made by the recommender system. Through these tasks, we focus on (1)~how scatterplots presented with different accuracy influence people's decisions and (2)~how scatterplots accompanied by accuracy presented in different formats impact trust.

Our findings show that scatterplots paired with accuracy do influence people's decisions and trust in recommendations. In addition, we found scatterplots accompanying descriptive accuracy (e.g., `high,' `low,' etc.) induced more trust. Finally, we observed that scatterplots not only assisted people in making decisions but also invalidating them. Taken together, the results provide new insights into how scatterplots assist in decision-making and trust in recommendations.

\section{Related Works}
\label{sec.background}

\textit{Trust and Transparency in Recommendations.}
It has been shown that people prefer human over algorithmic advice even if the algorithm performs better \cite{dietvorst2015algorithm}. Additionally, mistakes made by automated systems result in a quicker loss of confidence than those made by humans \cite{mahmud2022influences}. However, transparent and open explanations positively influence decision-making processes with machine learning \cite{lai2019human}, and precisely calibrating to transparency can significantly enhance user trust and the perceived quality of recommender systems \cite{kizilcec2016much,kunkel2019let}. Various studies have investigated the role of transparency in decisions with machine learning models \cite{yang2020visual,suresh2020misplaced,ribeiro2016should}, but, to the best of our knowledge, the role of visualization, particularly scatterplots, in providing transparency has not been thoroughly studied.

\textit{Visualizations in Machine Learning.}
Chatzimparmpas et al. present a state-of-the-art report on enhancing trust in machine learning models through interactive visualizations and outline research opportunities in their surveys \cite{chatzimparmpas2020state, chatzimparmpas2020survey}. Visualizations are crucial for deciphering black-box models by identifying key features and decoding complex behaviors. Tools such as the What If Tool \cite{wexler2019if} and SliceTeller \cite{zhang2022sliceteller} provide insights into model performance under various scenarios and identify data segments causing inaccuracies. Advances in interactive visualization techniques, such as RuleMatrix \cite{ming2018rulematrix}, EnsembleMatrix \cite{talbot2009ensemblematrix}, and deep learning interpretation systems \cite{kahng2017cti, wang2020visual, wu2020feature}, have improved the comprehension, exploration, and validation of predictive models. Tools such as ModelTracker \cite{amershi2015modeltracker} and Squares \cite{ren2016squares} refine error analysis and debugging processes through interactive visualizations that display summary statistics and instance-level performance metrics. Additionally, ExplainExplore \cite{collaris2020explainexplore} and Melody \cite{chan2020melody} analyze and summarize explanations through interactive systems for better understanding. Tools such as Fairkit-learn \cite{johnson2020fairkit}, FAIRVIS \cite{cabrera2019fairvis}, and FairSight \cite{ahn2019fairsight} help data scientists assess fairness, offering methods for subgroup analysis and fairness evaluation in algorithmic decisions. DiscriLens \cite{wang2020visual}, D-BIAS \cite{ghai2022d}, and Visual Auditor \cite{munechika2022visual} focus on uncovering bias, incorporating human insights, and providing analytical views on fairness issues. These efforts highlight a trend toward leveraging visual analytics for improved fairness and bias mitigation in machine learning, with less emphasis on improving general user trust in recommendations.

\textit{Visualization and Decision-Making.}
Visualizations guide users in making informed decisions \cite{savikhin2008applied, zhang2015designing}, communicate critical information \cite{dimara2018mitigating, savikhin2008applied}, aid in predictions with uncertainty \cite{guo2019visualizing}, and help identify complex patterns \cite{scheepens2015rationale}. Dimara et al. proposed an agenda to enhance decision-making through visualization research \cite{dimara2021critical}. Limited work employs visualizations to investigate trust in recommendation system decision-making. Xiong et al. used map-based visualizations to study transparency and trust through factors like accuracy, clarity, disclosure, and thoroughness \cite{xiong2019examining}. A study that closely relates to ours investigates how visual design impacts the perception of model bias, trust, and adoption \cite{10296507}. Findings reveal that visualization design choices, such as explaining fairness and mentioning bias, significantly affect trust levels. While both studies use visualizations and examine model performance metrics like accuracy, our study investigates the influence of scatterplots and accuracy on trust and decision-making, whereas their study focuses on identifying fair models using various signals and visualization methods, making these studies complementary.

\section{Study}
\label{sec.study}

We now describe our study of how scatterplots depicting recommender system predictions influence general users' trust and decision-making.

\textit{Procedure.}
The IRB-approved study was run on Amazon Mechanical Turk (AMT) using custom web pages and server built with Python.
First, participants received an overview of the study's goals and provided consent. After demographic questions, they completed tutorial questions with a sample dataset 
\begin{wrapfigure}{r}{0.49\linewidth}
    \centering

    \subfloat[\label{fig:Reference Scatterplot}]{
        \includegraphics[trim = 0pt 12pt 18pt 15pt, clip, height=1.675cm]{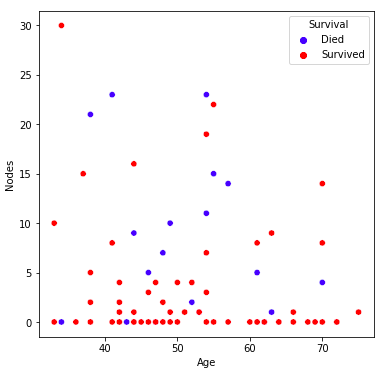}
    }
    \hfill    
    \subfloat[\label{fig:Recommender Scatterplot}]{
        {\includegraphics[trim = 0pt 12pt 20pt 15pt, clip, height=1.675cm]{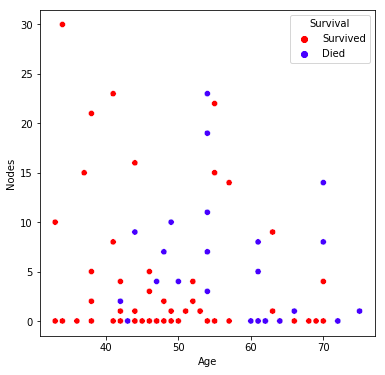}}
    }
    \hfill
    \subfloat[\label{fig:Single Recommendation Scatterplot}]{
        {\includegraphics[trim = 0pt 12pt 20pt 13pt, clip, height=1.675cm]{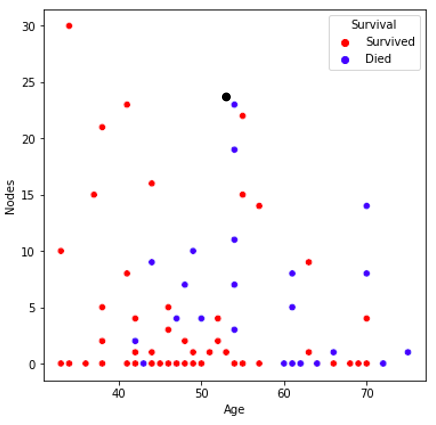}}
    }

    \caption{Experimental plots: (a)~reference plot of training data, (b)~recommender plot of testing data, and (c)~testing data with single recommendation in black.}
    \label{fig:Types of Scatterplots}

\end{wrapfigure}
to avoid learning effects. Participants then engaged in 2 types of task: (1)~high-level decision-making tasks (see \cref{sec.task1})
and (2)~low-level decision-making tasks (see \cref{sec.task2}).
An example experiment is at \supplement.

\begin{wrapfigure}{r}{0.61\linewidth}
    \centering
    
    \subfloat[Haberman\label{fig:Haberman's Survival Dataset}]{
        \includegraphics[trim = 0pt 12pt 18pt 15pt, clip, height=2.05cm]{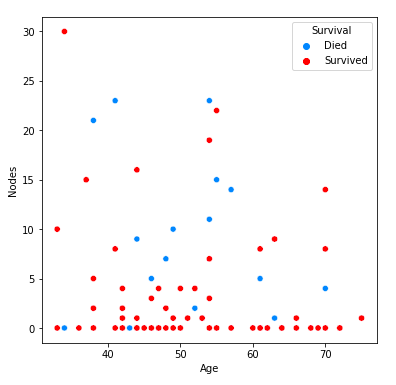}
    }
    \hfill
    \subfloat[Liver Disease\label{fig:Liver Disease Dataset}]{
        {\includegraphics[trim = 0pt 12pt 20pt 15pt, clip, height=2.05cm]{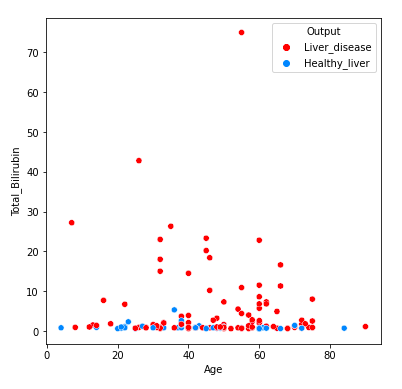}}
    }
    \hfill
    \subfloat[South German\label{fig:South German Credit Dataset}]{
        {\includegraphics[trim = 0pt 12pt 20pt 13pt, clip, height=2.05cm]{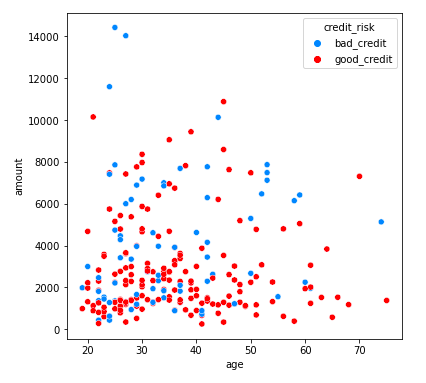}}
    }

    \caption{Datasets}
    \label{fig:Accuracy_Agreement}

\end{wrapfigure}

\textit{Visualizations.}
For all datasets, the 2 continuous attributes with the strongest correlation to the predicted value were selected to visualize. To reduce color bias, we generated the visualizations with 4 distinct color set orientations: Red/Blue, Blue/Red (see \cref{fig:Accuracy_Agreement}), Purple/Dogerblue, and Dogerblue/Purple (see \cref{fig:Numeric_Accuracy}).
Experiments used 3 types of scatterplots: (1)~reference scatterplots (i.e., ground truth from training data), where color represents the class data points actually belong to (see \cref{fig:Reference Scatterplot}); (2)~recommender scatterplots, where the color of the data points represents the class the system predicted they belong to (see \cref{fig:Recommender Scatterplot}); and
(3)~single recommendation scatterplots, where recommender scatterplots had an extra data point, depicted in black, to represent the specific data point for which the recommendation was being made (see \cref{fig:Single Recommendation Scatterplot}).

\textit{Recommender Models.}
We did not want to compare multiple models, but we did want to provide diversity in the recommendations. Therefore, we used 4 popular binary classification models: Decision Tree, Random Forest, K-Nearest Neighbors, and Support Vector Machine. 
As is the case with most real recommender systems, participants were not told which model they were being presented with during the experimental tasks.
All models were built by randomly partitioning each dataset into training and test 
sets at a 75\% to 25\% ratio, respectively. Each resulted in similar levels of accuracy that are reported at \supplement.

\textit{Data Sets.}
For our study, we used 3 datasets
acquired from the UCI Machine Learning Repository \cite{Dua:2019, groemping2019south}. The datasets were chosen to showcase some variety in data densities while avoiding complications of overdraw.
\textit{Haberman's Survival} (see \cref{fig:Haberman's Survival Dataset}) contains 306 instances of 3 attributes and deals with the survival of patients 5 years after undergoing surgery for cancer. \textit{Indian Liver Patient} (see 
\cref{fig:Liver Disease Dataset}) comprises 583 instances of 10 attributes about the health of individuals 
\begin{wrapfigure}{r}{0.43\linewidth}
  \centering
  \subfloat[Demographics\label{fig:age_gender}]{
 	\includegraphics[width=0.45\linewidth]{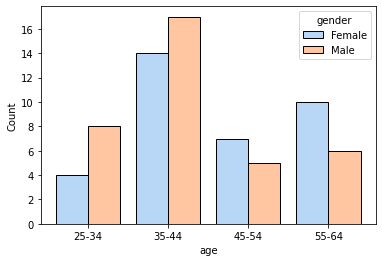}
    }
    \hfill
    \subfloat[Background\label{fig:visualization_ml}]{
  	\includegraphics[width=0.45\linewidth]{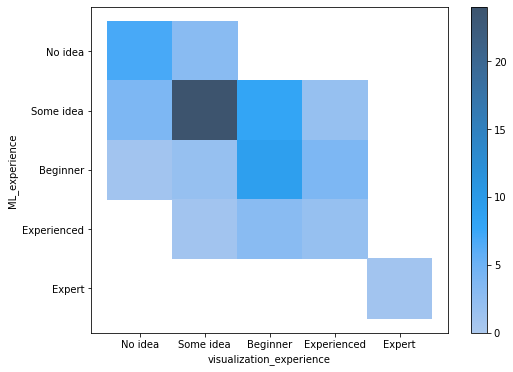}
    }
  \caption{(a)~Age and gender (female/blue and male/orange) and (b)~experience in machine learning (vert.) and visualization (horiz.).}
  \label{fig:demographics}
  
\end{wrapfigure}
with and without liver disease. \textit{South German Credit} (see \cref{fig:South German Credit Dataset}) consists of 1000 instances of 20 attributes with the credit quality of customers for a requested loan. Details of feature selection and feature visualizations are reported in the supplement (see \supplement).

\textit{Participants.}
120 subjects were paid \$2 to participate in the study, which lasted about 15 minutes. We screened the participants for their location and only those in the US and Canada were considered for the study.
We analyzed the data of 71 participants (35 female and 36 male) who passed the following inclusion criteria: (1)~answered all the experimental tasks without any omissions; and
(2)~provided a correct response to at least 1 of 2 attention-checks.  Demographics and self-reported experience with visualization and machine learning can be found in \cref{fig:demographics}.

\section{Task 1: Trust in Recommender Systems}
\label{sec.task1}

The primary aim of these tasks was to investigate the impact of scatterplots presented with the accuracy information on individuals' preferences for selecting a recommender system.

\subsection{Procedure}

\begin{wrapfigure}{r}{0.34\linewidth}
    \centering

    \includegraphics[trim=85pt 420pt 85pt 110pt, clip, width=0.97\linewidth]{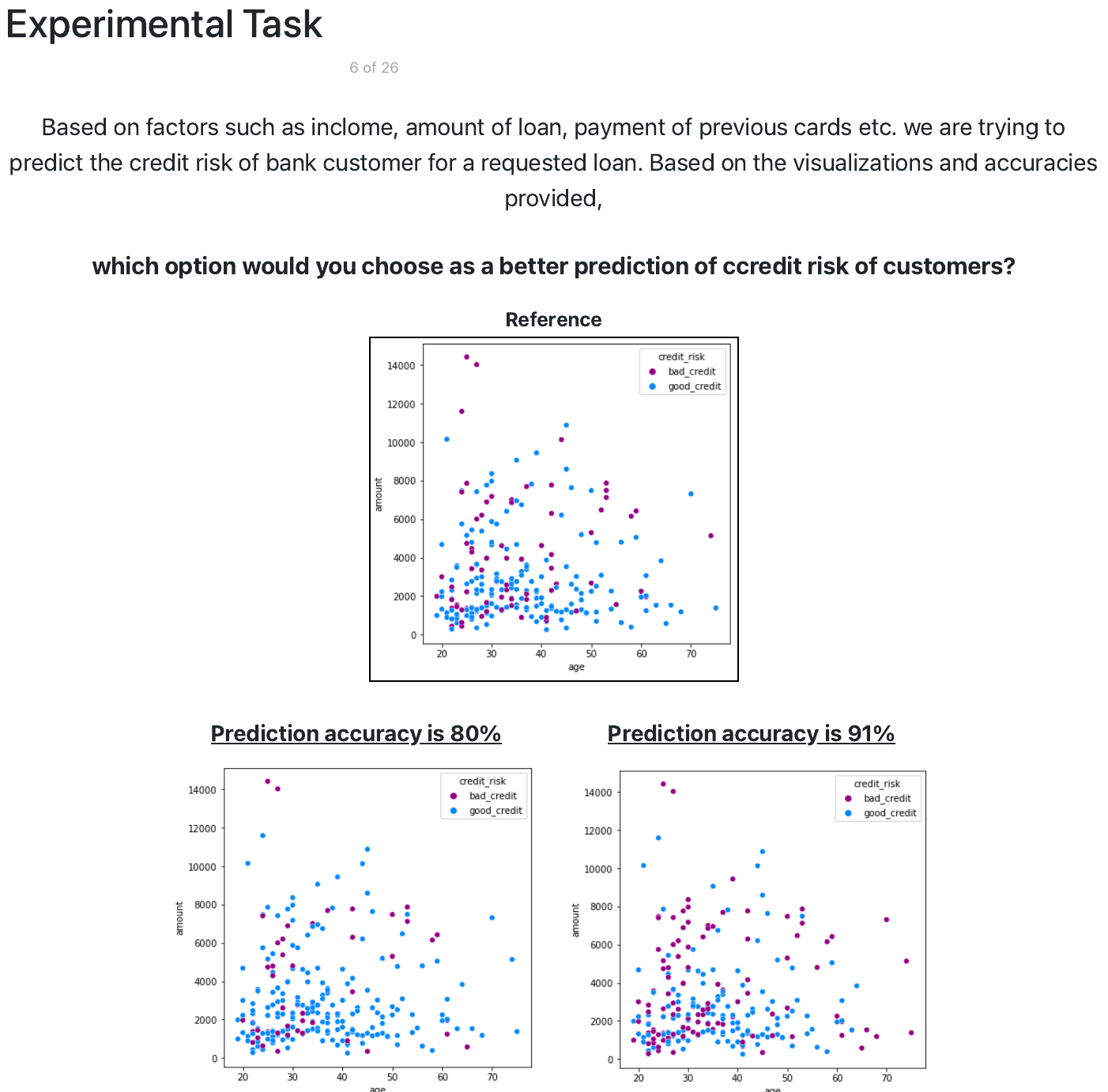}
    
    \caption{\textit{Visualization + Numeric Accuracy} version of the experiment 1 tasks.}
    \label{fig:Numeric_Accuracy}

\end{wrapfigure}

Each participant was presented with 3 visualizations: a reference scatterplot and 2 recommender scatterplots (see \cref{sec.study}). Their objective was to select a recommender system from the 2 available options. We designed 3 versions of the task. Each participant was presented with all 3 versions, making it a within-subject stimuli. The variations were:
\begin{itemize}
    \item The \textit{Visualization-Only} version presented a reference scatterplot and 2 recommender scatterplot visualizations (example in supplement).
    \item The \textit{Visualization + Numeric Accuracy} version presented a reference scatterplot with 2 recommender scatterplots, each accompanied by numeric accuracy (e.g., \textit{91\%}) for each of the recommender scatterplots (see \cref{fig:Numeric_Accuracy}). 
    \item The \textit{Visualization + Descriptive Accuracy} version followed a similar setup, with a reference scatterplot and 2 recommender scatterplots, each accompanied by descriptive accuracy labels (e.g., \textit{high}, \textit{medium}, or \textit{low}), instead of numeric accuracy values (example in supplement).
\end{itemize}

\textit{Recommender Systems Used.} 
We employed all 4 recommender system models (see \cref{sec.study}). For every question, 2 models were chosen at random.

\textit{Accuracy.}
In the real-world, accuracy is often provided without significant context. We intentionally avoided providing participants with an explanation of 
what the accuracy actually meant to determine better how participants naturally perceived it. 
Further, we opted to ignore the model output accuracy and instead systematically generated an accuracy for each stimulus.  For binary classifiers like those we used, 50\% accuracy indicates that the algorithm is essentially guessing. 
Using this value as a lower bound, we categorized accuracy into 3 tiers, aligning descriptive accuracy with numeric accuracy: \textit{high} $\rightarrow$ $[90\%,95\%]$; \textit{medium} $\rightarrow$ $[75\%,80\%]$; and \textit{low} $\rightarrow$ $[65\%,70\%]$.
The precise numeric accuracy displayed for each question was chosen at random from within the designated range.

\textit{Color.} Participants were randomly assigned to 1 of the 4 distinct color treatments (see \cref{sec.study}), and all of their stimuli used that color combination.

\textit{Tasks.} Each participant was assigned 9 tasks in random order, encompassing the 3 experimental conditions (i.e., Visualization-Only, Visualization + Numeric Accuracy, and Visualization + Descriptive Accuracy) across the 3 datasets (i.e., Haberman's Survival, Indian Liver Patient, and South German Credit).

\subsection{Results}

We first evaluated whether there was any potential association (i.e., bias) for recommender system selection and task versions. We conducted a 2-way ANOVA that showed 
no statistically significant interaction
(F(6,24) = 2.43, p=.057),
suggesting that participants did not have a preference for any specific recommender model but rather made decisions based on the information provided.

\noindent
\begin{minipage}[t]{\linewidth}
    \textbf{Hypothesis 1 [H1]:} \textit{Scatterplots accompanied by higher accuracy correspond to a greater chance of selection compared to those accompanied by lower accuracy.}
\end{minipage}

\begin{wrapfigure}{r}{0.51\linewidth}

  \centering
  \subfloat[High, Medium, and Low Accuracy\label{fig:H1}]{
 	\includegraphics[width=0.42\linewidth]{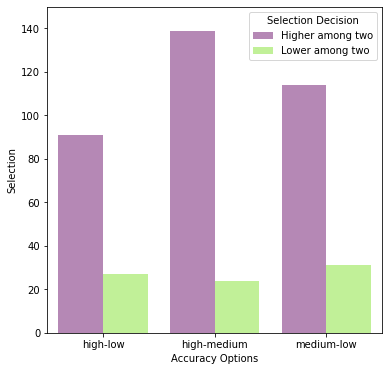}
    }
    \hfill
    \subfloat[Numeric and Descriptive Accuracy\label{fig:H2}]{
  	\includegraphics[width=0.42\linewidth]{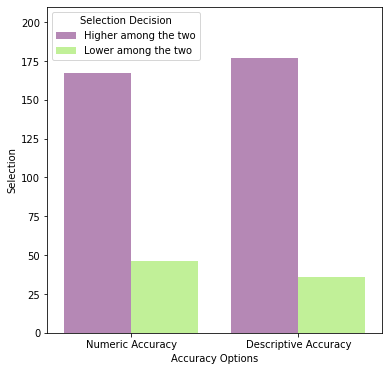}
    }
  \caption{Selection of recommender systems: (a)~presents the proportion of time the recommender with higher accuracy was selected, while (b)~presents the selection of recommender systems for numeric and descriptive accuracy conditions.}
  \label{fig:Hypothesis-task1}

\end{wrapfigure}

We investigated the influence of scatterplots accompanied by higher accuracy (both numeric and descriptive) on the selection of recommender systems, as compared to scatterplots accompanied by lower accuracy. We used Shapiro-Wilk test to check for normality, and the results indicated the data deviated from a normal distribution (W=0.813, p<.001). Thus, we conducted a Kruskal-Wallis non-parametric test, and the results of the test indicate that \textit{scatterplot visualizations presented with higher accuracy significantly impact the selection of recommender systems} (H(1, n=142)=161.19, p<.001). For the post hoc analysis, we performed a Dunn's test and compared the mean ranks. The results indicated a statistical significance for the selection of scatterplots that are accompanied by higher accuracy (Z(1, n=142)=12.69, p<.001). The data was visualized to show the effects (see \cref{fig:H1}). These findings are consistent with our hypothesis, leading us to reject the null hypothesis and \textbf{accept H1}.

\noindent
\begin{center}
\begin{minipage}[t]{\linewidth}
\textbf{Hypothesis 2 [H2]:} \textit{Scatterplots accompanied by numeric accuracy will have less influence on selection than those accompanied by descriptive accuracy.}
\end{minipage}
\end{center}

For the subsequent analysis, we examined whether the method of presenting accuracy, specifically Numeric Accuracy (e.g., \textit{90\%}) versus Descriptive Accuracy (e.g., \textit{high}), alongside scatterplot visualizations influenced the selection of recommender systems. 
We used the Shapiro-Wilk test to check for normality, and the results indicated the data deviated from a normal distribution (W=0.698, p<.001). Due to the non-normal distribution of data, we conducted a Wilcoxon test. The results revealed that the \textit{method of presenting accuracy metrics alongside scatterplot visualizations did not exhibit a statistically significant association with the selection of recommender systems} (W=276, p=.189) (see \cref{fig:H2}).  Therefore, we are unable to reject the null hypothesis and, therefore, \textbf{reject H2}.

\textit{Discussion.} \textbf{H1} validated that accuracy influences trust and decisions regarding the recommender system. On the other hand, we expected that the ambiguity associated with numeric accuracy would cause participants to ``trust the visualization'' less, as opposed to the less ambiguous descriptive accuracy. However, as \textbf{H2} was rejected, we cannot support that claim in the situation.

\section{Task 2: Trust in Individual Recommendations}
\label{sec.task2}

Building on the previous task, this task focuses on examining the way scatterplots, paired with accuracy, impact participants' trust in individual recommendations generated by the recommender systems.

\subsection{Procedure}
Participants were presented with information about the dataset, including what the recommender system is recommending in a single recommendation scatterplot (see \cref{fig:Single Recommendation Scatterplot}) coupled with the recommender systems' recommendation for the data point. They were then required to indicate if they would agree with the recommendation. The decision to agree with the recommendation was considered a proxy for trust in the recommendation. 
 Additionally, in order to investigate the behavior of the participant's decision when additional information was given, we altered the sequence in which the information was provided.

\begin{wrapfigure}{r}{0.37\linewidth}
    
    \begin{minipage}[m]{0.87\linewidth}
    \subfloat[First part of the task\label{fig:NA2_SGC_V11}]{
        \includegraphics[trim=85pt 550pt 65pt 110pt, clip, width=\linewidth]{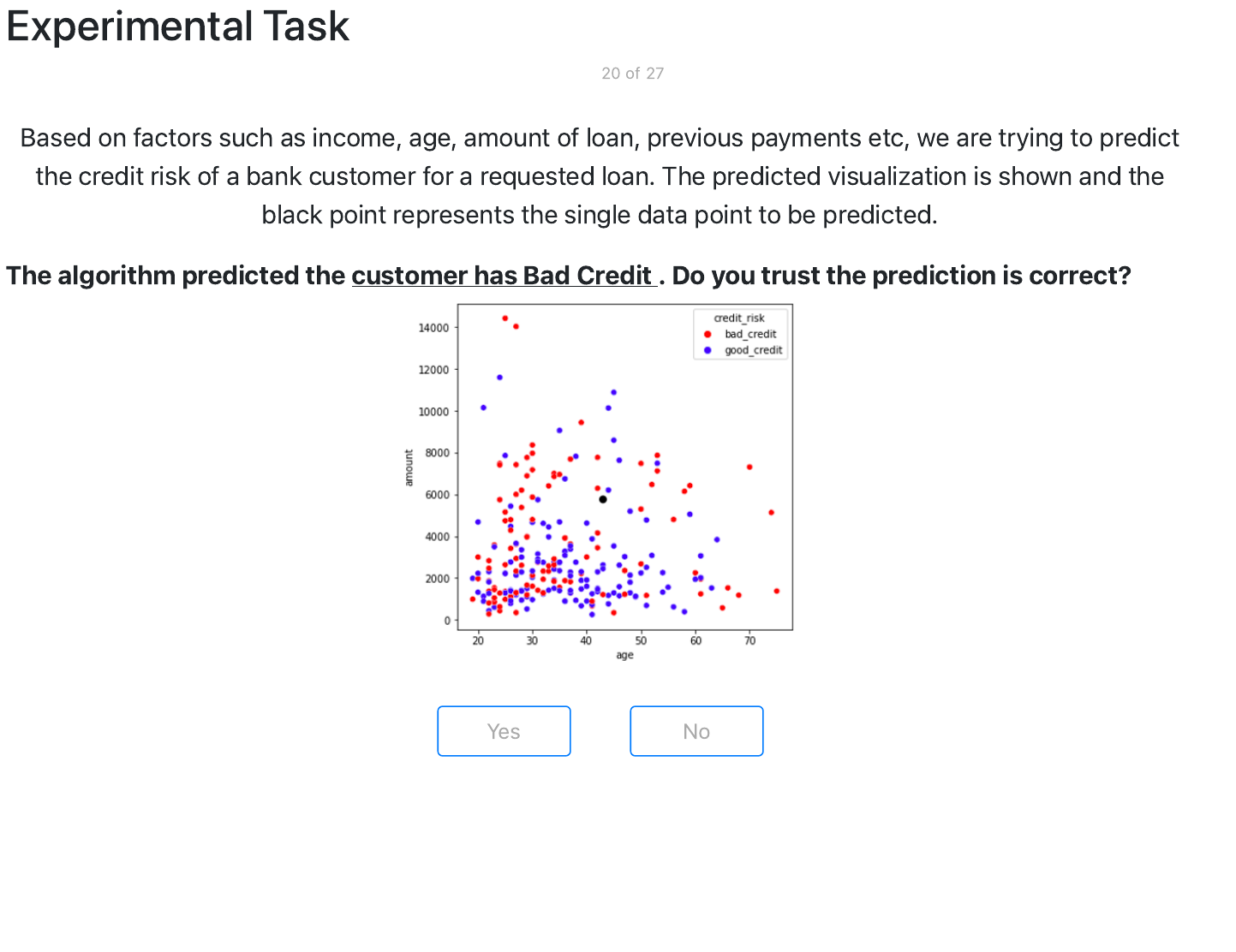}
    }
    \end{minipage}
    \begin{minipage}[m]{0.87\linewidth}
    \subfloat[Second part of the task\label{fig:NA2_SGC_V12}]{
  	\includegraphics[trim=85pt 550pt 65pt 110pt, clip, width=\linewidth]{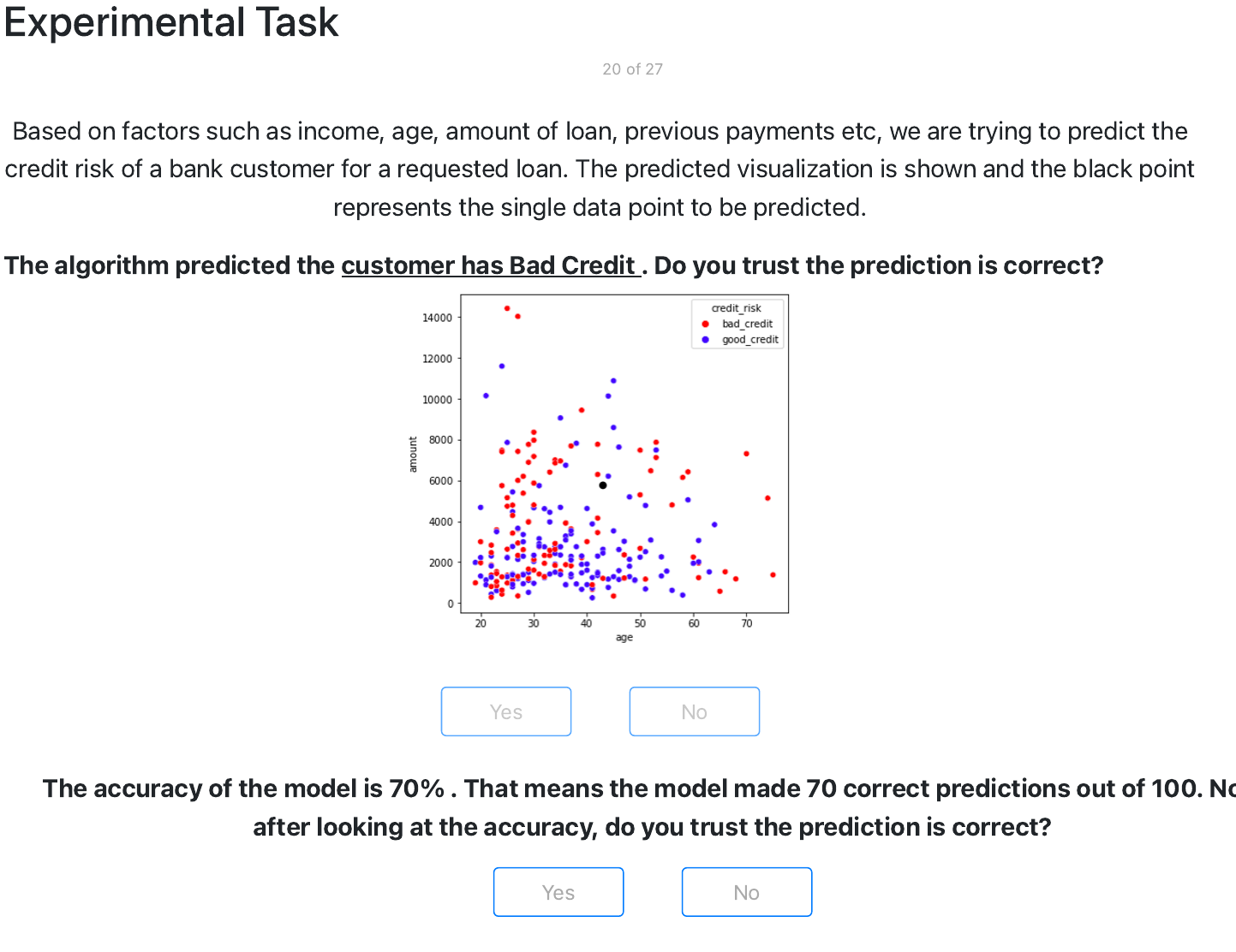}
    }
    \end{minipage}

    \caption{\textit{Visualization $\rightarrow$ Numeric Accuracy} version of task 2. (a)~illustrates the initial decision of the task, while (b)~the second step provides an opportunity to reconsider their decision.  }
    \label{fig:Accuracy_Visualization_class}

\end{wrapfigure}

\begin{itemize}
    \item The \textit{Visualization + Accuracy} condition provided all information at once and asked for agreement with the recommendation (example in supplement). 
    \item The \textit{Visualization $\rightarrow$ Accuracy} condition first showed the recommendation and the single recommendation scatterplot and asked for agreement with the recommendation (see \cref{fig:NA2_SGC_V11}). Subsequently, accuracy information was provided, and participants were asked if they still agreed with the recommendation (see \cref{fig:NA2_SGC_V12}). 
    \item The \textit{Accuracy $\rightarrow$ Visualization} condition initially presented the recommendation and the accuracy and asked for agreement with it. Subsequently, the single recommendation scatterplot was shown, and participants were asked if they still agreed with the recommendation (example in supplement).
\end{itemize}

Participants were randomly assigned to 1 of 12 distinct treatments in a \textit{4 * 3} design. As with Task~1, participants were randomly assigned to 1 of the 4 distinct color treatments (see \cref{sec.study}). All of their stimuli used that color combination, making it a between-subject stimuli. The task type (i.e., \textit{Visualization + Accuracy}, \textit{Visualization $\rightarrow$ Accuracy}, and \textit{Accuracy $\rightarrow$ Visualization}) was the second between subject stimuli. On the other hand, participants were presented with both the numeric (e.g., \textit{91\%}) and descriptive (e.g., \textit{high}) conditions, making it a within-subject stimuli. Other variables, such as recommender systems' recommendation and the accuracy shown, were randomly assigned to each participant for each task.

\textit{Task.} Participants were given information about the dataset, including what the recommendation system is recommending, a single recommendation scatterplot, and the systems' recommendation of the data point and the accuracy.  They were asked to record their agreement with the recommendation by selecting either \textit{Yes} or \textit{No} (see \cref{fig:NA2_SGC_V11}). For the \textit{Visualization $\rightarrow$ Accuracy} and \textit{Accuracy $\rightarrow$ Visualization} conditions, they were then shown additional information and asked the same question again (see \cref{fig:NA2_SGC_V12}). Each participant responded to a total of 6 tasks---3 datasets and 2 types of accuracy.

\textit{Recommender Systems and Recommendations.} We employed all 4 recommenders (see \cref{sec.study}). 3 data points, excluded from both training and test sets, were chosen from each dataset for recommendation. All 4 recommender systems yielded identical recommendations for the data points. We presented participants with either the actual recommendations made by the system or the inverted recommendations to balance the behavior toward stimuli and avoid bias. The recommendations were randomly assigned to participants, meaning while a participant was presented with the actual recommendation for 1 task, they might encounter the incorrect recommendation for the subsequent task. 

\textit{Accuracy.} To assess the impact of varying accuracy, we ignored the output accuracy and systematically generated accuracy similar to Task 1.

\subsection{Results}

\noindent
\begin{center}
\begin{minipage}[t]{\linewidth}
\textbf{Hypothesis 3 [H3]}: \textit{Scatterplots accompanied by higher accuracy correspond to a greater chance of trusting the recommendation compared to scatterplots accompanied by lower accuracy.}
\end{minipage}
\end{center}

\begin{wrapfigure}{r}{0.485\linewidth}

  \centering
  \subfloat[Level of accuracy\label{fig:H3}]{
 	\includegraphics[width=0.45\linewidth]{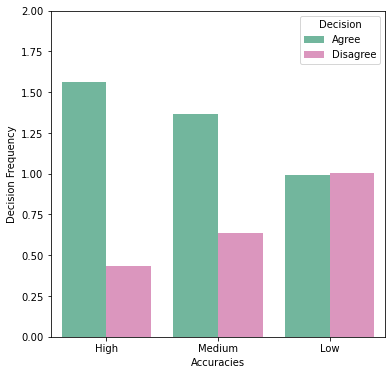}
    }
    \hfill
    \subfloat[Num. or Desc.\label{fig:H4}]{
  	\includegraphics[width=0.45\linewidth]{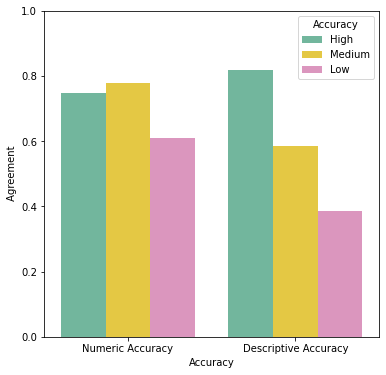}
    }
  \caption{Agreement with the recommendation: (a)~across high, medium, and low accuracy and (b)~when accuracy is presented as Numeric or Descriptive.}
  \label{fig:Hypothesis-task2}

\end{wrapfigure}

We filtered the full data to test this hypothesis. Our initial analysis involved examining if the data satisfy the assumptions of a 1-way ANOVA. The results from the Shapiro-Wilk test (W=0.665, p<0.001) indicated that data deviated significantly from a normal distribution. This led us to choose non parametric Kruskal-Wallis test. The Kruskal-Wallis analysis (H(2, n=280)=15.352, p<.001) indicated the difference was statistically significant. This supported that \textit{scatterplots accompanied by higher accuracy correspond to a greater chance of building trust in the recommendation compared to scatterplots accompanied by lower accuracy}. The visualization of the corresponding data also supports this finding (see \cref{fig:H3}). This led us to reject the null hypothesis and \textbf{accept H3}.

\noindent
\begin{center}
\begin{minipage}[t]{\linewidth}
\textbf{Hypothesis 4 [H4]}: \textit{Scatterplots accompanied by Descriptive Accuracy correspond to a greater chance of trusting the recommendation compared to scatterplots accompanied by Numeric Accuracy.}
\end{minipage}
\end{center}

We again filtered the data needed to test this hypothesis from the full data. As each participant was exposed to both scatterplots along with Numeric Accuracy and scatterplots along with Descriptive Accuracy an equal number of times, we decided to use a Paired Samples T-test. The Shapiro-Wilk test was used to test data normality, and the results (W=0.931, p<.001) indicated the distribution of data deviated from normality. We selected a non parametric alternative, Wilcoxon sign test on paired samples to test our hypothesis. The results indicated (V=419, p=.008) that the median \textit{trust of scatterplots accompanied by Numeric Accuracy was less than that of scatterplots accompanied by Descriptive Accuracy}. The visualizations from the data also reflected a similar relation (see \cref{fig:H4}). This led us to reject null hypothesis and \textbf{accept H4}.

\noindent
\begin{minipage}[t]{\linewidth}
\textbf{Hypothesis 5 [H5]}: \textit{The order of presentation of scatterplots and accuracy of information influence the participant's trust in the recommendation.}
\end{minipage}

We utilized data for \textit{Visualization + Accuracy}, \textit{Visualization $\rightarrow$ Accuracy} and \textit{Accuracy $\rightarrow$ Visualization} stimuli. 
We used Shapiro-Wilk to test for normality, and the results (W=0.852, p<.001) implied data deviated from a normal distribution. Therefore, we avoided 1-way ANOVA and instead used a non-parametric alternative, Kruskal-Wallis test. The Kruskal-Wallis test results (H (2, n=142)=1.005, p=.604) implied that \textit{no strong evidence exists to suggest that the order of presentation of scatterplots and accuracy influenced decisions}. Based on the results, we could not reject the null hypothesis and therefore \textbf{reject H5}.

\textit{Discussion.}  In this task, \textbf{H3} showed that scatterplots accompanied by higher accuracy assist in making trustworthy decisions about a recommendation. Furthermore, unlike the previous experimental task, \textbf{H4} validated that scatterplots accompanied by subjective information like Descriptive Accuracy aid in instilling trust in a recommendation compared to Numeric Accuracy. Although we expected the order of presenting information to sway the trust in recommendations, we did not find evidence of it in our experiment.

\section{Qualitative Tasks}
\label{sec.qualitative}

\begin{wrapfigure}{r}{0.43\linewidth}
  \centering

  \subfloat[Dependency levels\label{fig:likert_image}]{\hspace{7pt}\includegraphics[width=0.49\linewidth]{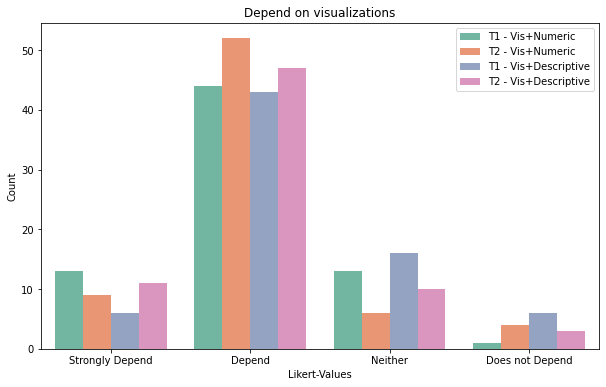}\hspace{7pt}}
  \hfill
  \subfloat[Strategies\label{fig:open_image}]{\includegraphics[width=0.40\linewidth]{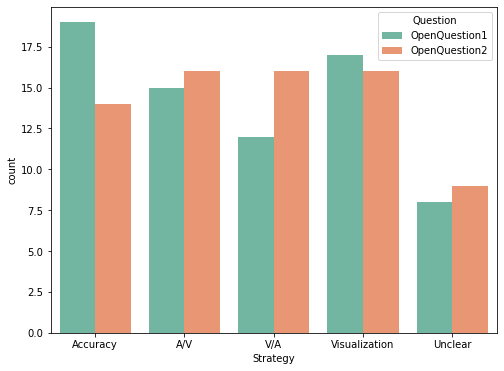}}

  \caption{(a)~Self-reported dependence on scatterplots for task 1 (T1) and task 2 (T2). (b)~The outcomes derived from the post-experiment questionnaire, where \texttt{OpenQuestion1} was for recommender systems, and \texttt{OpenQuestion2} was for individual recommendations (AV: Accuracy-Validation, VA: Visualization-Validation).}
  \label{fig:open}

\end{wrapfigure}

We evaluated participants' self-reported dependence on scatterplots and examine the decision-making strategies they used.

\textit{Role of Scatterplots.}
At the end of each experimental task version, participants recorded their level of dependence on scatterplots using a 5-point Likert Scale ranging from `Strongly Depend' to `Strongly Does not Depend'. 
Our analysis revealed that over 75\% of participants indicated a significant level of dependence on the visualization, with many reporting either `Strongly Depend'  or `Depend' on scatterplots for decision-making (see \cref{fig:likert_image}).

\textit{Post Experiment Questionnaire.}
After the experimental tasks, participants were asked 2 open-ended questions. 
The first prompted them to record the approach employed in deciding to pick a recommender system when a scatterplot was coupled with accuracy (\texttt{OpenQuestion1}), while the
second asked about the approach utilized in determining whether or not to trust an individual recommendation (\texttt{OpenQuestion2}). %
The responses were reviewed and organized into 5 categories based on their decision strategies. 
(1)~\textit{Visualization-Based} included participants who expressed that decisions were based primarily on scatterplots. %
(2)~\textit{Accuracy-Based} was assigned to participants who indicated accuracy as the primary determinant of their decisions. 
(3)~\textit{Visualization-Validation} was when the decisions were initially made using scatterplots and subsequently validated based on accuracy. 
(4)~\textit{Accuracy-Validation} first relied on accuracy to make decisions and then validated the decisions based on scatterplots. 
(5)~Participants whose decision-making strategy lacked specificity, often involving factors such as gut feeling and instinct, were categorized as \textit{Unclear}.
The results suggest that when faced with the choice between 2 recommender systems, participants tended to give slightly higher priority to accuracy over scatterplots (see~\cref{fig:open_image}). However, our analysis revealed that when making a decision regarding their trust in individual recommendations, scatterplots played a slightly more significant role in influencing decisions.

\section{Discussion and Conclusion}
\label{sec.discussion}

\textit{Discussion.}
We analyzed the way scatterplots presented, along with accuracy, influence trust in recommender systems. We observed that participants relied on the accuracy provided with the scatterplot to make decisions, and higher accuracy positively influenced both deciding on selecting a recommender system and trusting an individual recommendation. %
While the presentation of accuracy accompanying scatterplots as numeric and descriptive did not show a statistically significant difference in selecting a recommender system, Descriptive Accuracy exhibited a notable impact on trust in individual recommendations.
We believe the lack of influence on the selection of a recommender system could be due to the multiway comparison of the scatterplot and accuracy of both recommender systems.
Moreover, the findings also revealed that participants utilized scatterplots to validate their decisions when accuracy was higher, but they particularly relied on scatterplots to make decisions as accuracy values dropped.

\textit{Limitations and Future Work.}
While we primarily focused on scatterplot visualizations in our study, we plan to extend our research to examine the influence of additional visualization types on trust and decision-making. We visualized the relation between 2 attributes and aimed to explore how visualizing multiple attributes affects people's decisions on recommendations.
We recognize that real-world datasets are often large, which can clutter visualizations and become difficult to interpret. 
Finally, we acknowledge the subjective nature of trust and the challenges in measuring it. In our work, we primarily consider relative trust by focusing on participants' selection of one option over another.

\textit{Conclusion.} The inspiration behind this research was to explore and comprehend the role of visualizations, especially scatterplots presented alongside accuracy, in influencing trust in recommender systems and individual recommendations. Our observations show participants did depend on scatterplots for making decisions and validating them. In particular, participants tended to trust high accuracy alone, but visualizations became increasingly more important with lower accuracy. We also saw that the presentation of the accuracy in a descriptive form, although vague, was easier to interpret than the numeric form, which also has contextualized meaning. 
Based on our findings, we advocate for incorporating visualizations to enhance trust and decision-making in recommender systems. This approach helps mitigate the issues that arise from relying solely on accuracy metrics, thereby increasing the ability of the general population to make more informed decisions.

\bibliographystyle{splncs04}
\bibliography{main}

\begin{thebibliography}{10}
\providecommand{\url}[1]{\texttt{#1}}
\providecommand{\urlprefix}{URL }
\providecommand{\doi}[1]{https://doi.org/#1}

\bibitem{ahn2019fairsight}
Ahn, Y., Lin, Y.R.: Fairsight: Visual analytics for fairness in decision
  making. IEEE TVCG  \textbf{26}(1),  1086--1095 (2019)

\bibitem{amershi2015modeltracker}
Amershi, S., Chickering, M., Drucker, S.M., Lee, B., Simard, P., Suh, J.:
  Modeltracker: Redesigning performance analysis tools for machine learning.
  In: ACM CHI. pp. 337--346 (2015)

\bibitem{cabrera2019fairvis}
Cabrera, {\'A}., Epperson, W., et~al.: Fairvis: Visual analytics for
  discovering intersectional bias in machine learning. In: IEEE VAST. pp.
  46--56 (2019)

\bibitem{chan2020melody}
Chan, G.Y.Y., Bertini, E., Nonato, L.G., Barr, B., Silva, C.T.: Melody:
  Generating and visualizing machine learning model summary to understand data
  and classifiers together. arXiv  (2020)

\bibitem{chatzimparmpas2020survey}
Chatzimparmpas, A., Martins, R.M., Jusufi, I., Kerren, A.: A survey of surveys
  on the use of visualization for interpreting machine learning models.
  Information Visualization  \textbf{19}(3),  207--233 (2020)

\bibitem{chatzimparmpas2020state}
Chatzimparmpas, A., Martins, R.M., Jusufi, I., Kucher, K., Rossi, F., Kerren,
  A.: The state of the art in enhancing trust in machine learning models with
  the use of visualizations. Computer Graphics Forum  \textbf{39}(3),  713--756
  (2020)

\bibitem{chouldechova2018case}
Chouldechova, A., Benavides-Prado, D., Fialko, O., Vaithianathan, R.: A case
  study of algorithm-assisted decision making in child maltreatment hotline
  screening decisions. In: ACM FAccT. pp. 134--148 (2018)

\bibitem{collaris2020explainexplore}
Collaris, D., van Wijk, J.J.: Explainexplore: Visual exploration of machine
  learning explanations. In: IEEE PacificVis. pp. 26--35 (2020)

\bibitem{dietvorst2015algorithm}
Dietvorst, B.J., Simmons, J.P., Massey, C.: Algorithm aversion: people
  erroneously avoid algorithms after seeing them err. J.\ Exp.\ Psych.: General
   \textbf{144}(1), ~114 (2015)

\bibitem{dimara2018mitigating}
Dimara, E., Bailly, G., Bezerianos, A., Franconeri, S.: Mitigating the
  attraction effect with visualizations. IEEE TVCG  \textbf{25}(1),  850--860
  (2018)

\bibitem{dimara2021critical}
Dimara, E., Stasko, J.: A critical reflection on visualization research: Where
  do decision making tasks hide? IEEE TVCG  \textbf{28}(1),  1128--1138 (2021)

\bibitem{Dua:2019}
Dua, D., Graff, C.: {UCI} machine learning repository (2017)

\bibitem{surendran2020movie}
Furtado, F., Singh, A.: Movie recommendation system using machine learning
  algorithms. International Journal of Research in Engineering and Technology
  (2020)

\bibitem{10296507}
Gaba, A., Kaufman, Z., Cheung, J., Shvakel, M., Hall, K.W., Brun, Y.,
  Bearfield, C.X.: My model is unfair, do people even care? visual design
  affects trust and perceived bias in machine learning. IEEE TVCG
  \textbf{30}(1),  327--337 (2024)

\bibitem{ghai2022d}
Ghai, B., Mueller, K.: D-bias: a causality-based human-in-the-loop system for
  tackling algorithmic bias. IEEE TVCG  \textbf{29}(1),  473--482 (2022)

\bibitem{groemping2019south}
Groemping, U.: South german credit data: Correcting a widely used data set.
  Rep. Math., Phys. Chem., Berlin, Germany, Tech. Rep  \textbf{4}, ~2019 (2019)

\bibitem{guo2019visualizing}
Guo, S., Du, F., Malik, S., Koh, E., Kim, S., Liu, Z., Kim, D., Zha, H., Cao,
  N.: Visualizing uncertainty and alternatives in event sequence predictions.
  In: ACM CHI. pp. 1--12 (2019)

\bibitem{johnson2020fairkit}
Johnson, B., Bartola, J., Angell, R., Witty, S., Giguere, S., Brun, Y.:
  Fairkit, fairkit, on the wall, who’s the fairest of them all? supporting
  fairness-related decision-making. EURO Journal on Decision Processes
  \textbf{11},  100031 (2023)

\bibitem{kahng2017cti}
Kahng, M., Andrews, P.Y., Kalro, A., Chau, D.H.: Activis: Visual exploration of
  industry-scale deep neural network models. IEEE TVCG  \textbf{24}(1),  88--97
  (2017)

\bibitem{kay2016ish}
Kay, M., Kola, T., Hullman, J.R., Munson, S.A.: When (ish) is my bus?
  user-centered visualizations of uncertainty in everyday, mobile predictive
  systems. In: ACM CHI. pp. 5092--5103 (2016)

\bibitem{kizilcec2016much}
Kizilcec, R.F.: How much information? effects of transparency on trust in an
  algorithmic interface. In: ACM CHI. pp. 2390--2395 (2016)

\bibitem{kube2019allocating}
Kube, A., Das, S., Fowler, P.J.: Allocating interventions based on predicted
  outcomes: A case study on homelessness services. AAAI  \textbf{33}(01),
  622--629 (2019)

\bibitem{kunkel2019let}
Kunkel, J., Donkers, T., et~al.: Let me explain: Impact of personal and
  impersonal explanations on trust in recommender systems. In: ACM CHI. pp.
  1--12 (2019)

\bibitem{lai2019human}
Lai, V., Tan, C.: On human predictions with explanations and predictions of
  machine learning models. In: ACM FAccT. pp. 29--38 (2019)

\bibitem{mahmud2022influences}
Mahmud, H., Islam, A.N., Ahmed, S.I., Smolander, K.: What influences
  algorithmic decision-making? a systematic literature review on algorithm
  aversion. Technological Forecasting and Social Change  \textbf{175},  121390
  (2022)

\bibitem{ming2018rulematrix}
Ming, Y., Qu, H., Bertini, E.: Rulematrix: Visualizing and understanding
  classifiers with rules. IEEE TVCG  \textbf{25}(1),  342--352 (2018)

\bibitem{munechika2022visual}
Munechika, D., Wang, Z., Reidy, J., et~al.: Visual auditor: Interactive
  visualization for detection and summarization of model biases. In: IEEE VIS.
  pp. 45--49 (2022)

\bibitem{ren2016squares}
Ren, D., Amershi, S., Lee, B., Suh, J., Williams, J.D.: Squares: Supporting
  interactive performance analysis for multiclass classifiers. IEEE TVCG
  (2016)

\bibitem{ribeiro2016should}
Ribeiro, M.T., Singh, S., Guestrin, C.: " why should i trust you?" explaining
  the predictions of any classifier. In: ACM KDD. pp. 1135--1144 (2016)

\bibitem{savikhin2008applied}
Savikhin, A., Maciejewski, R., Ebert, D.S.: Applied visual analytics for
  economic decision-making. In: IEEE VAST. pp. 107--114 (2008)

\bibitem{scheepens2015rationale}
Scheepens, R., Michels, S., van~de Wetering, H., van Wijk, J.J.: Rationale
  visualization for safety and security. In: Computer Graphics Forum. pp.
  191--200 (2015)

\bibitem{suresh2020misplaced}
Suresh, H., Lao, N., Liccardi, I.: Misplaced trust: Measuring the interference
  of machine learning in human decision-making. In: ACM Web Science (2020)

\bibitem{talbot2009ensemblematrix}
Talbot, J., Lee, B., et~al.: Ensemblematrix: interactive visualization to
  support machine learning with multiple classifiers. In: ACM CHI. pp.
  1283--1292 (2009)

\bibitem{wang2020visual}
Wang, Q., Xu, Z., Chen, Z., Wang, Y., Liu, S., Qu, H.: Visual analysis of
  discrimination in machine learning. IEEE TVCG  \textbf{27}(2),  1470--1480
  (2020)

\bibitem{wexler2019if}
Wexler, J., Pushkarna, M., Bolukbasi, T., Wattenberg, M., et~al.: The what-if
  tool: Interactive probing of machine learning models. IEEE TVCG
  \textbf{26}(1),  56--65 (2019)

\bibitem{wu2020feature}
Wu, C., Qian, A., et~al.: Feature-oriented design of visual analytics system
  for interpretable deep learning based intrusion detection. In: TASE. pp.
  73--80 (2020)

\bibitem{xiong2019examining}
Xiong, C., Padilla, L., Grayson, K., Franconeri, S.: Examining the components
  of trust in map-based visualizations. In: TrustVis Workshop. pp. 19--23
  (2019)

\bibitem{yang2020visual}
Yang, F., Huang, Z., Scholtz, J., Arendt, D.L.: How do visual explanations
  foster end users' appropriate trust in machine learning? In: ACM IUI. pp.
  189--201 (2020)

\bibitem{yin2019understanding}
Yin, M., Wortman~Vaughan, J., Wallach, H.: Understanding the effect of accuracy
  on trust in machine learning models. In: ACM CHI. pp. 1--12 (2019)

\bibitem{zhang2022sliceteller}
Zhang, X., Ono, J.P., Song, H., Gou, L., et~al.: Sliceteller: A data
  slice-driven approach for machine learning model validation. IEEE TVCG
  \textbf{29}(1),  842--852 (2022)

\bibitem{zhang2015designing}
Zhang, Y., Bellamy, R.K., Kellogg, W.A.: Designing information for remediating
  cognitive biases in decision-making. In: ACM CHI. pp. 2211--2220 (2015)

\end{thebibliography}

\end{document}